\documentclass[12pt]{article}
\usepackage[dvips]{graphicx}
\usepackage{epsfig}
\usepackage{amsmath,amssymb,amsbsy}
\usepackage{amsfonts}
\usepackage{bm}
\newcommand{\be}{\begin{equation}}
\newcommand{\ee}{\end{equation}}
\newcommand{\beq}{\begin{eqnarray}}
\newcommand{\eeq}{\end{eqnarray}}

\begin{document}
\thispagestyle{empty}
\begin{center}

{\Large\bf{Do we understand elastic scattering up to LHC energies? \footnote{Invited talk at " DIFFRACTION 2012"'
Puerto del Carmen, Lanzarote Canary Islands (Spain), September 10-15 2012, to appear in
AIP Conference Proceedings (2013)}}}

\vskip 0.5cm
{\bf Jacques Soffer}
\vskip 0.3cm
Physics Department, Temple University,\\
Barton Hall, 1900 N, 13th Street\\
Philadelphia, PA 19122-6082, USA
\vskip 0.5cm
\title{Do we understand elastic scattering up to LHC energies?}
\end{center}
{\bf Abstract}\\
The measurements of high energy  $\bar p p~\mbox{and}~p p$ elastic at ISR, SPS, and Tevatron colliders
have provided usefull informations on the behavior of the scattering
amplitude. A large step in energy
domain is accomplished with the LHC collider presently running, giving a unique
opportunity to improve our knowledge on the asymptotic regime of the elastic scattering
amplitude and to verify the validity of our theoretical approach, to describe the total cross section
$\sigma_{tot}(s)$, the total elastic cross section $\sigma_{el}(s)$, the ratio of the real to imaginary parts of the forward amplitude $\rho(s)$ and the differential cross section 
${d\sigma (s,t)/dt}$.

 \vspace*{3mm}

\vspace*{ 2cm}
The Bourrely-Soffer-Wu (BSW) model was first proposed, in 1978 \cite{bsw}, to describe the experimental data on elastic $p p$ and $\bar{p} p$, taken at the relatively low energies available
to experiments, more than forty years ago. Some more complete analysis were done later \cite{bsw1},  showing very successful theoretical predictions for these processes, at earlier colliders. Since a new energy domain is now accessible with the LHC collider at CERN \cite{deile}, it is a good time to recall the main features of the BSW model and to check its validity. The spin-independent elastic scattering amplitude is given by
\begin{equation}
  \label{eq:one}
  a(s,t) = \frac{is}{2\pi}\int e^{-i\mathbf{q}\cdot\mathbf{b}} (1 - 
e^{-\Omega_0(s,\mathbf{b})})  d\mathbf{b} \ ,
\end{equation}
where $\mathbf q$ is the momentum transfer ($t={-\bf q}^2$) and 
$\Omega_0(s,\mathbf{b})$ is the opaqueness at impact parameter 
$\mathbf b$ and at a given energy $s$, the square of the center-of-mass energy. We take the simple form
%\begin{equation}
%\label{eq:omega}
$\Omega_0(s,\mathbf{b}) = S_0(s)F(\mathbf{b}^2)+ R_0(s,\mathbf{b})$, 
%\end{equation}
the first term is associated with the "Pomeron" exchange, which generates 
the diffractive component of the scattering and the second term is 
the Regge background which is negligible at high energy.
The function $S_0(s)$ is given by the complex symmetric expression, obtained from the high energy behavior
of quantum field theory \cite{chengWu}
\begin{equation}
  \label{eq:Sdef}
  S_0(s)= \frac{s^c}{(\ln s)^{c'}} + \frac{u^c}{(\ln u)^{c'}}~,
\end{equation}
with $s$ and $u$ in units of $\mbox{GeV}^2$, where $u$ is the
third Mandelstam variable. In Eq. (\ref{eq:Sdef}), $c$ and $c'$ are two 
dimensionless constants given above \footnote{In the Abelian case one finds $c' = 3/2$ and it was
conjectured that in Yang-Mills non-Abelian gauge theory
one would get $c' = 3/4$ (T.T. Wu private communication).}
in Table 1. That they
are constants implies that the Pomeron is a fixed Regge cut, rather than a
Regge pole.\\ For the asymptotic behavior at high energy and modest momentum
transfers, we have to a good approximation
so that
\begin{equation}
  \label{eq:S2}
    S_0(s)= \frac{s^c}{(\ln s)^{c'}} + 
\frac{s^ce^{-i\pi c}}{(\ln s-i\pi)^{c'}}~.
\end{equation}
The choice one makes for $F(\mathbf{b}^2)$ is essential and we take the Bessel transform of
\begin{equation}
  \label{eq:FtildDef}
  \tilde{F}(t) = f[G(t)]^2\frac{a^2+t}{a^2-t}~,
\end{equation}
where $G(t)$ stands for the proton "` nuclear form factor"', parametrized similarly to the electromagnetic
form factor, with two poles $G(t)=1/(1- t/m_1^2)(1- t /m_2^2)$.
The remaining four parameters of the model, $f$, $a$, $m_1$ $\mbox{and}$ $m_2$, are given in Table 1. We
 define the ratio of the real to imaginary parts of the forward amplitude 
$\rho(s) = \frac{\mbox{Re}~a(s, t=0)}{\mbox{Im}~a(s, t=0)}$,
 the total cross section 
$\sigma_{tot} (s) = (4\pi/s)\mbox{Im}~a(s, t=0)$, 
the differential cross section
${d\sigma (s,t)/dt} = \frac{\pi}{s^2}|a(s,t)|^2$, 
and the integrated elastic cross section 
$\sigma_{el} (s) = \int dt \frac{d\sigma (s,t)} { dt}$.
 \begin{table}
    \centering
        \caption{\label{tab:table1} Parameters of the BSW model \cite{bsw1}.}
        \vspace{0.4cm}
    \begin{tabular}{|rllrll|}\hline
$c$ & = & 0.167, & \;\;$c'$&=&0.748\\
$m_1$&=&0.577 GeV,&\;\;$m_2$&=&1.719 GeV\\
$a$&=&1.858 GeV,&\;\;${\it f}$&=&6.971 GeV$^{-2}$\\\hline
    \end{tabular}
      \end{table}

%\vspace*{-20mm}
\begin{figure}[hbt]
\begin{center}
%\rule{5cm}{0.2mm}\hfill\rule{5cm}{0.2mm}
%\vskip 2.5cm
%\rule{5cm}{0.2mm}\hfill\rule{5cm}{0.2mm}
\psfig{figure=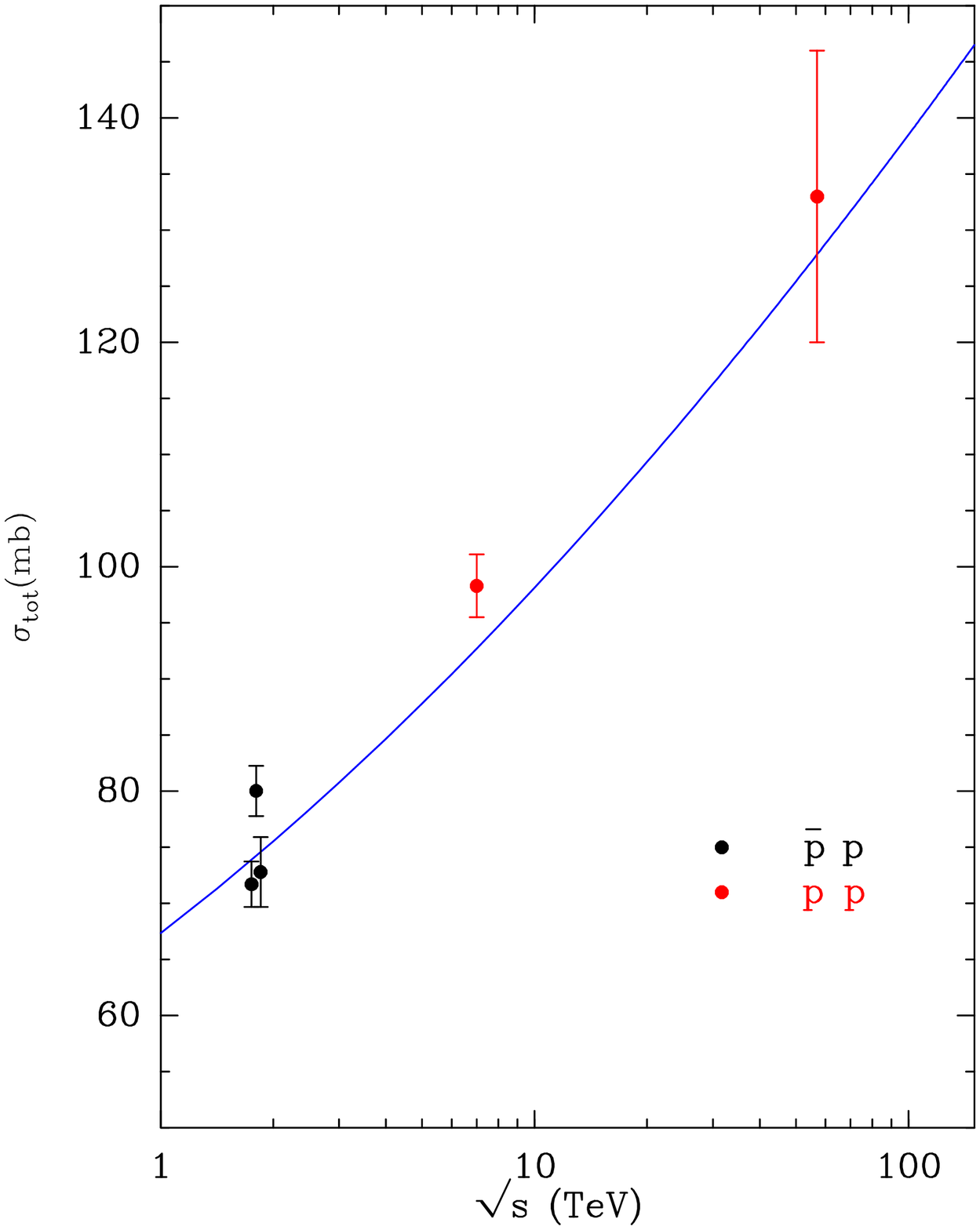,height=3.0in}
\hskip 1cm
\psfig{figure=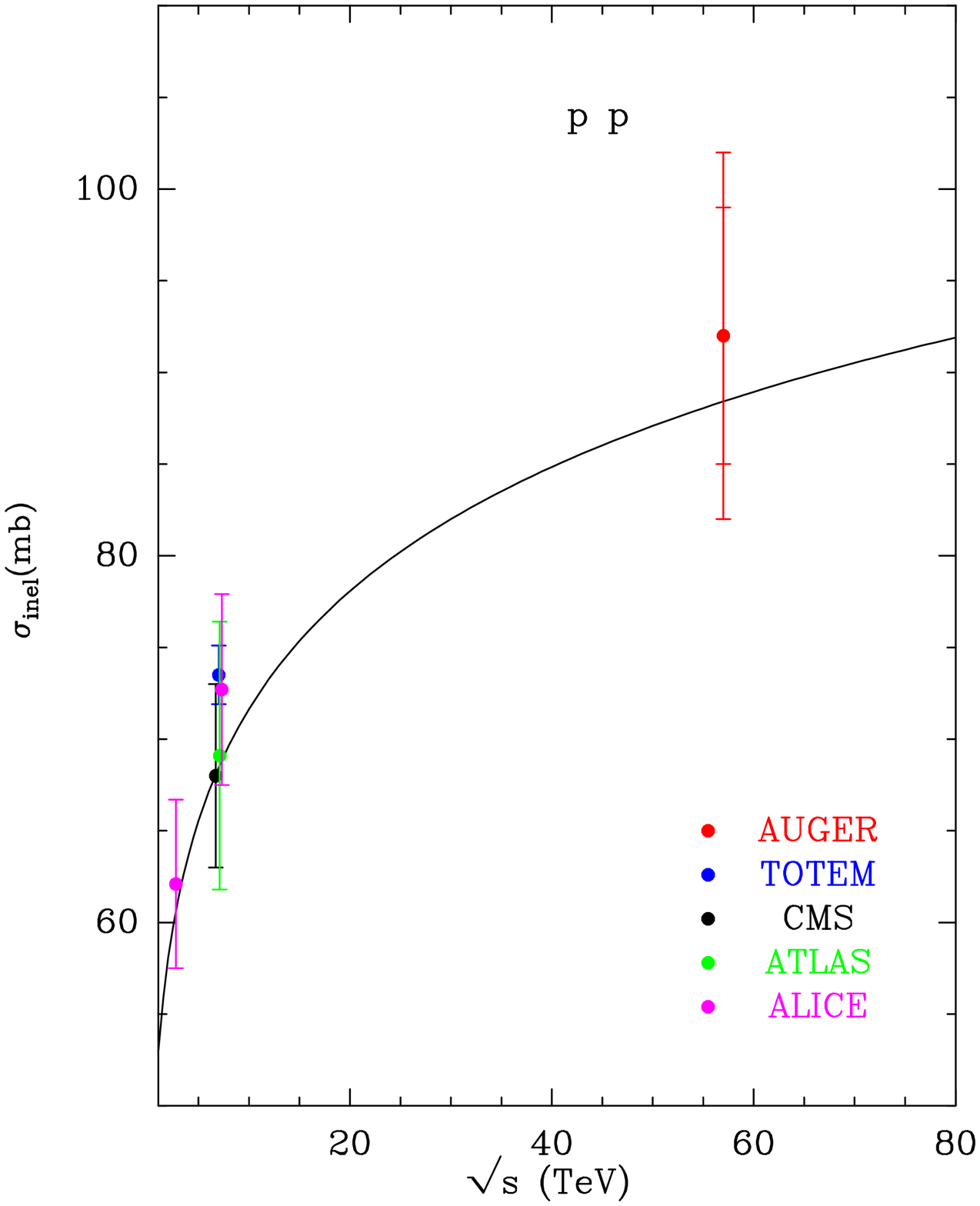,height=3.0in}
\end{center}
\vspace*{-8mm}
\caption{ $pp$ ($\bar pp$) elastic scattering, $\sigma_{tot}$, ({\it Left}), $\sigma_{inel}$ ({\it Right}) as a function of the energy. The solid curves are the BSW predictions}.
\label{fig:sig-totel}
\end{figure}
One important feature of the BSW model is, as a consequence of Eq. (\ref{eq:S2}), the fact that the phase
of the amplitude is built in. Therefore real and imaginary parts of the amplitude cannot be chosen independently and we will
now see how to test them, according to different $t$ regions.\\
Let us first consider the total cross section which is directly related to $\mbox{Im}~a(s, t=0)$. We
show in Fig. \ref{fig:sig-totel}
 (Left) our prediction up to cosmic rays energy.
 The BSW approach predicts at 7 TeV $\sigma_{tot}= 93.6\pm 1 \mbox{mb}$. Two other important quantities are the integrated elastic cross section $\sigma_{el}$, which is predicted to be $\sigma_{el}= 24.8 \pm 0.3\mbox{mb}$ and finally the total inelastic cross section defined as $\sigma_{inel}=\sigma_{tot}-\sigma_{el}$.\\
These predictions must be compared with different new experimental LHC results \cite{deile}, namely, from TOTEM, 
$\sigma_{tot}= (98.0 \pm 2.5)\mbox{mb}$, $\sigma_{el}= 24.8 \pm 0.2(stat) \pm 1.2(syst)\mbox{mb}$ and
$\sigma_{inel}= 73.5 \pm 0.6(stat) + 1.8(-1.3)(syst) \mbox{mb}$, from ATLAS which has found $\sigma_{inel}= 69.4 \pm 2.4(expt) \pm 6.9(extra) \mbox{mb}$ and from CMS, which has reported $\sigma_{inel}= 68 \pm 2(syst) \pm 2.4(lum) \pm 4(extra) \mbox{mb}$.  We show in Fig. \ref{fig:sig-totel}
 (Right) a compilation of $\sigma_{inel}$, up to cosmic rays energy.
We notice that our $\sigma_{inel}$ is in excellent agreement with the last two determinations, but although our $\sigma_{el}$ agrees very well with the value of TOTEM, our prediction for $\sigma_{tot}$ is lower but consistent with their value.\\
Another specific feature of the BSW model is the fact that it incorporates the theory of expanding protons \cite{chengWu}, with the 
physical consequence that the ratio $\sigma_{el}/\sigma_{tot}$ increases with energy. This is precisely in agreement with the LHC data \cite{deile} and when $s \to \infty$ one expects $\sigma_{el}/\sigma_{tot}\to 1/2$, which is the black disk limit.\\
The BSW model predicts the correct real part of the forward elastic amplitude $\rho(s)$, which appears to have a flat energy dependence in the high energy region and in the black disk limit $s \to \infty$, one expects $\rho(s)\to 0$.
The importance of the value of $\rho$ at the LHC has been emphasized in Ref.\cite{bkmst} and we are glad to report that this measurement is now in progress near the very forward direction, to reach $|t|\sim 6 \cdot10^{-4}\mbox{GeV}^2$.\\
 Before moving to the non-forward region let us mention another test of the BSW amplitude, with the analyzing power $A_N$, near the very forward direction. In this kinematic region, the so called Coulomb nuclear interference (CNI) region, $A_N$ results from the interference of the Coulomb amplitude which is purely real, with the imaginary part of the hadronic non-flip amplitude, namely $a(s, t)$, if one assumes that there is no contribution from the single-flip hadronic amplitude \cite{bsw4}. New data \cite{star} confirm a zero single-flip hadronic amplitude and the right determination of $\mbox{Im}~a(s, t)$ in the CNI region.\\
The non-forward region allows us to understand the behavior of the differential cross section from the $t$-dependence of the real and imaginary parts of the scattering amplitude, which have both some zeros at different $t$ values, as shown in Fig. 2 (Left). The imaginary part dominates over the real part, except when the imaginary part has a zero, producing either a shoulder, for ${\bar pp}$ at $\sqrt{s}=1.8\mbox{GeV}$ around $|t|=0.6 \mbox{GeV}^2$, or a real dip for $pp$ at $\sqrt{s}=7\mbox{GeV}$, as in Fig. 2 (Right) around $|t|=0.5 \mbox{GeV}^2$. Our prediction is in excellent agreement with the Tevatron data \cite{bsw5,d0} but although we predict the right position of the dip at LHC, we seem to underestimate the forward slope and to overestimate the cross section in the region of the second maximum, determined by TOTEM \cite{deile}.
\begin{figure}[hbt]
\begin{center}
  \begin{minipage}{6.5cm}
  \epsfig{figure=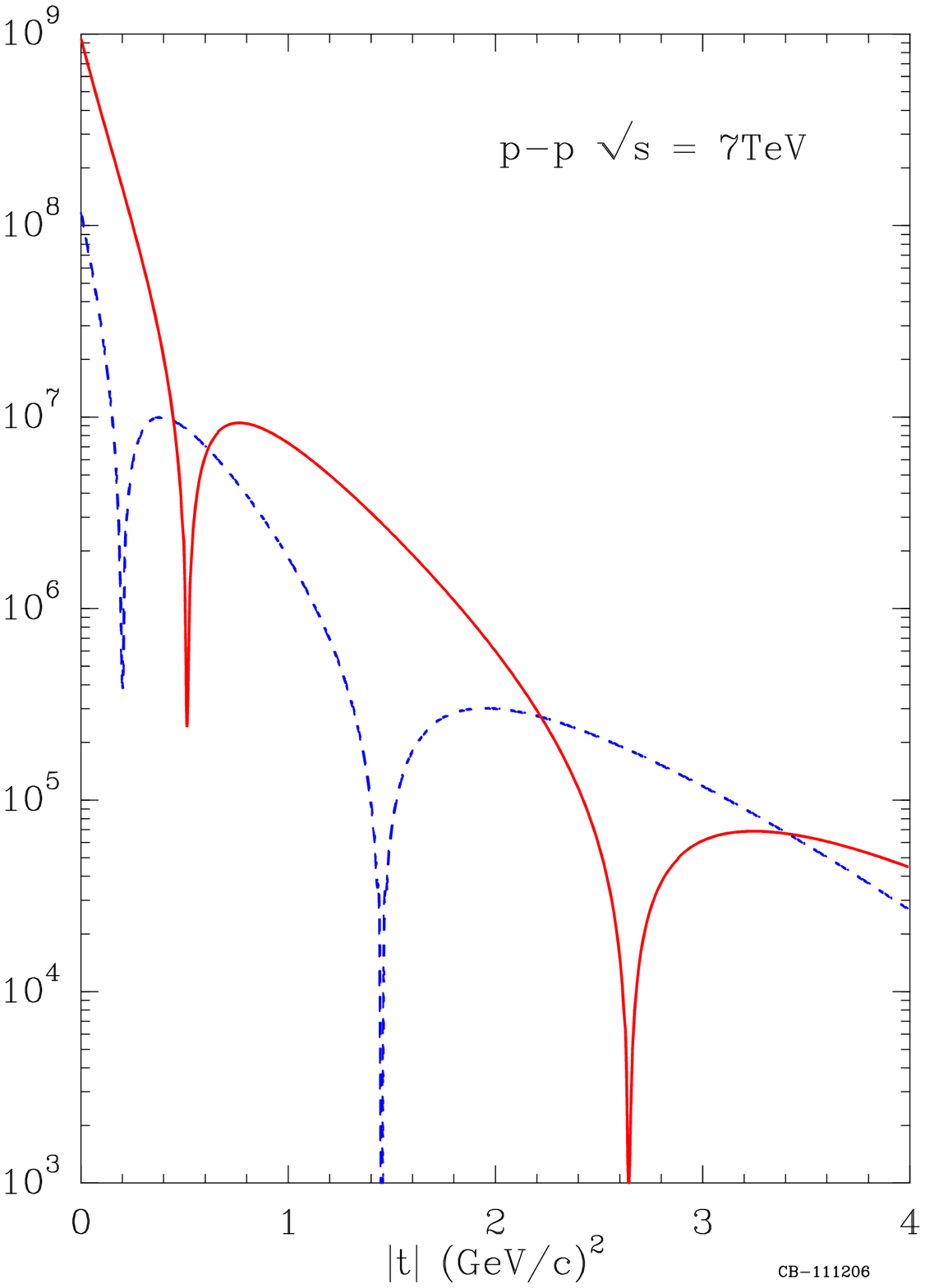,width=6.8cm}
  \end{minipage}
    \begin{minipage}{6.5cm}
  \epsfig{figure=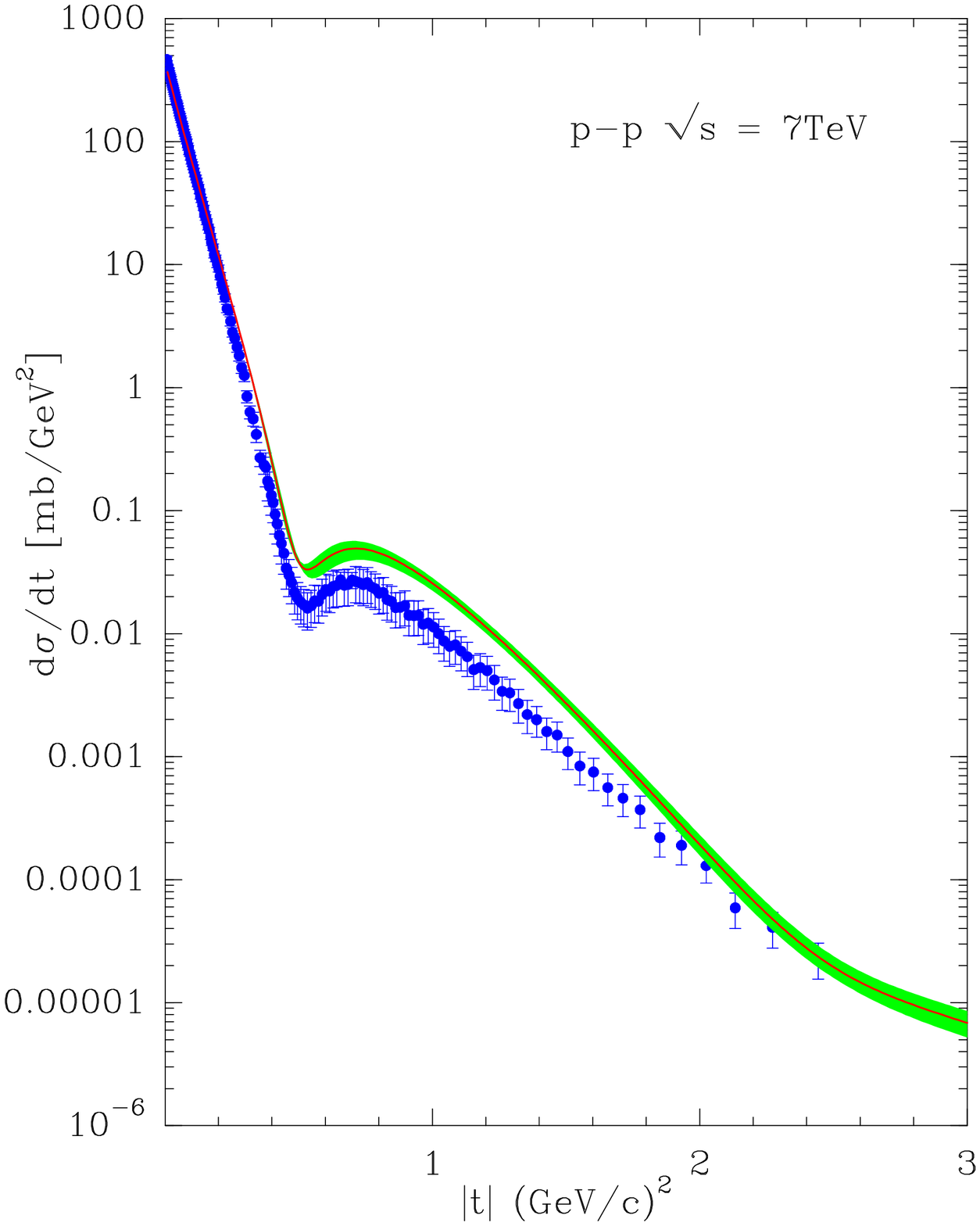,width=7.2cm}
    \end{minipage}
\end{center}
  \vspace*{-5mm}
\caption{{\it (Left)}: The absolute value of the $pp$ elastic scattering amplitude, {\it solid curve} $|\mbox{Im}~a(s,t)|$, {\it dashed curve} $|\mbox{Re}~a(s,t)|,$ versus $t$ at $\sqrt{s}=7\mbox{TeV}$. ({\it Right}): The corresponding differential cross section (The curves are taken from Ref.\cite{bsw6}) and the LHC data is from Ref.\cite{deile}).}
\label{fi:fig2}
\vspace*{-1.5ex}
\end{figure}

LHC is opening up a new area for $pp$ elastic scattering and 
TOTEM has confirmed the following basic features expected at LHC from BSW:
 $\sigma_{tot}$ and $\sigma_{el}/\sigma_{tot}$ increase,
 the diffraction peak is still shrinking, the dip position is moving in and the second maximum is moving up.
 So far one observes only partial quantitative agreement with the BSW approach, but more data are needed, in particular from ATLAS-ALFA, which should be released soon hopefully. We also look forward to a precise measurement of $\rho$.
%%%%%%%%%%%%%%%%%%%%%%%%%%%%%%%%%%%%%%%%%%%%

\end{document}